\documentclass[reprint,superscriptaddress,frontmatterverbose,amsmath,amssymb,aps,prd,floatfix,nofootinbib]{revtex4-2}

\usepackage[T1]{fontenc}
\usepackage{lmodern}
\usepackage[english]{babel}
\usepackage{graphicx}
\usepackage{dcolumn}
\usepackage{bm}
\usepackage{hyperref}
\usepackage{orcidlink}
\usepackage{booktabs}
\usepackage{multirow}
\usepackage{float}
\usepackage[caption=false]{subfig}
\usepackage{soul}

\usepackage[activate={true,nocompatibility},final,
tracking=true,kerning=true,spacing=true,
factor=1100,stretch=10]{microtype}
\SetTracking{encoding={*}, shape=sc}{40}
\microtypecontext{spacing=nonfrench}

\usepackage[dvipsnames]{xcolor}
\definecolor{blue}{RGB}{0,0,128}

\usepackage{hyperref}
\hypersetup{colorlinks,allcolors=blue,pageanchor=false}

\def \beq{\begin{equation}}
\def \eeq{\end{equation}}
\def \bea{\begin{eqnarray}}
\def \eea{\end{eqnarray}}
\def \bma{\begin{matrix}}
\def \ema{\end{matrix}}

\def \({\left(}
\def \){\right)}
\def \[{\left[}
\def \]{\right]}
\def \nn{\nonumber}
\def \nl{\nn\\}

\def \cL{{\cal L}}

\def \Z2{\mathbb{Z}_2}

\def \La{\Lambda}

\NewDocumentCommand{\pewcp}{}{P^{\prime C}_{\mathrm{EW}}}
\NewDocumentCommand{\pewp}{}{P^{\prime}_{\mathrm{EW}}}

\NewDocumentCommand{\gn}{}{g_N}

\NewDocumentCommand{\BtoKinv}{}{B^+ \to K^+ + \mathrm{inv}}
\NewDocumentCommand{\KLtopinv}{}{K_L \to \pi^0 + \mathrm{inv}}

\begin{document}
\allowdisplaybreaks

\begin{flushright}
{\footnotesize
USC-TH-2026-03
}
\end{flushright}%

\title{Axion-like particles and sterile neutrinos solve the $B\to K\nu\bar\nu$ and $B\to \pi K$ puzzles}

\author{Bhubanjyoti Bhattacharya\,\orcidlink{0000-0003-2238-321X}}
\email{bbhattach@ltu.edu}
\affiliation{Department of Natural Sciences, Lawrence Technological University, Southfield, MI 48075, USA}
\author{Alakabha Datta\,\orcidlink{0000-0001-8713-2783}}
\email{datta@phy.olemiss.edu}
\affiliation{Department of Physics and Astronomy, University of Mississippi, Oxford, MS 38677, USA}
\author{Girish Kumar\,\orcidlink{0000-0001-6051-2495}}
\email{girish89@sc.edu}
\affiliation{Department of Physics and Astronomy, University of South Carolina, Columbia, SC 29208, USA}
\author{Danny Marfatia}
\email{dmarf8@hawaii.edu}
\affiliation{Department of Physics and Astronomy, University of Hawaii, Honolulu, HI 96822, USA}


\begin{abstract}
  The recent measurement of the branching ratio of $B^+ \to K^+ + \mathrm{inv}$ (where ``inv'' denotes invisible states) by the Belle~II
  collaboration is enhanced relative to the standard model  expectation by 2.7$\sigma$. An older puzzle persists in measurements of the branching ratios and CP asymmetries of
  $B \to \pi K $ decays. We address these two
  anomalies in flavor-changing neutral current $B$ decays, with a short-lived axion-like
  particle (ALP) with mass close to that of the $\pi^0$. In the model with the minimum number of new couplings, the ALP has
  couplings to the photon, top quark and a heavy sterile neutrino.  The ALP contributes to the
  $ B \to \pi^0 K $ decays by mixing with the $\pi^0$.  It contributes to
  $B^+ \to K^+ + \mathrm{inv}$ by its off-shell coupling to sterile neutrino pairs. The model can explain the excess in the total rate, but not the observed distribution of signal events. We make predictions for all $B \to K^{(*)} + \mathrm{inv}$ modes and for the rare kaon decays, $K^+ \to \pi^+ + \mathrm{inv}$ and
  $K_L \to \pi^0 + \mathrm{inv}$. We find an appreciable contribution to the magnetic moment of the
  muon, and negligible contributions to the magnetic moment of the electron and $b \to s e^+ e^-$.
\end{abstract}

\maketitle

\section{Introduction}\label{sec:intro}
There is general consensus that the standard model (SM) is incomplete. The flavor structure of any new interactions may be revealed in current and upcoming high-precision experiments probing the intensity frontier. New physics may consist of very heavy particles (whose effects may be described by higher-dimensional operators) or of light and weakly coupled particles.

Some well-known models of light, weakly coupled new physics have axion or axionlike particles (ALP), or light sterile neutrinos.  Here, we consider an ALP and its effects in various meson decays. We employ a model-independent description of the ALP Lagrangian, and the main ALP decay modes are to invisible final states or to two photons via its mixing with pseudoscalar mesons. Recent results on $B$ meson decays to invisible final states by the Belle~II experiment~\cite{Belle-II:2023esi} indicate an enhanced rate relative to its SM expectation and provides motivation to search for ALP signals in this decay.

The long-standing $B\to\pi K$ puzzle refers to the tension between SM predictions and experimental measurements of the four $B\to\pi K$ decay branching ratios and direct CP asymmetries. Global fits to the data~\cite{
pdg,LHCb:2020byh,Belle-II:2023ksq,Belle-II:2023grc} reveal a persistent tension with the SM, particularly in direct CP-violating asymmetries in $B^+ \to \pi^0 K^+$ and $ B^0 \to \pi^- K^+$ decays where the predicted near-equality is contradicted by experimentally measured values of opposite signs. 
ALP mixing with the pseudoscalar states $\pi^0,\eta$, and $\eta'$ can lead to deviations from SM expectations in hadronic decays. 
An application of this mechanism to explain the $B\to\pi K$ puzzle was discussed in Ref.~\cite{Bhattacharya:2021shk}, while a solution to the puzzle with a long-lived axion close to the pion mass was considered in Ref.~\cite{Altmannshofer:2024kxb}. In this paper, we study the interplay of ALP interactions in several interesting experimental signals in flavor physics. The first such signal is the Belle~II collaboration's measurement of the branching ratio for $B$ decay to a kaon and invisible final states~\cite{Belle-II:2023esi}:
\begin{eqnarray}
  {\cal B}(B^+\to K^+ + {\rm inv}) &=& (2.3\pm0.5({\rm stat})^{+0.5}_{-0.4}({\rm syst})) \nl && \hspace{2truecm}\times10^{-5}\,.
  \label{BelleIIanomaly}
\end{eqnarray}
In the SM, this decay is dominated by $B^+\to K^+\nu{\bar\nu}$. The measured branching ratio for $B^+\to K^+ + {\rm inv}$ exceeds the SM expectation for ${\cal B}(B^+\to K^+\nu{\bar\nu})$ from the HPQCD collaboration by $\sim 2.7\sigma$~\cite{Parrott:2022zte}.\footnote{Note that the HPQCD calculation uses $|V_{tb}V_{ts}^\ast|=0.04185(93)$~\cite{Parrott:2022zte}. This is close to the value obtained from CKM unitarity using the inclusive determination of $|V_{cb}| = (41.97 \pm 0.48) \times 10^{-3}$~\cite{Finauri:2023kte,Fael:2024rys}. Using instead the smaller exclusive value of $|V_{cb}| = (39.62\pm 0.47) \times 10^{-3}$~\cite{HeavyFlavorAveragingGroupHFLAV:2024ctg,Vittorio:2025txp} lowers the SM prediction by about $9\%$, slightly increasing the discrepancy with the Belle~II measurement. } The presence of invisible particles in the final state makes this process a very powerful probe of new particles that couple weakly to the SM.

The NA62 collaboration recently reported an analogous measurement in the $K$ system. Based on data collected between 2016 and 2024 they find~\cite{NA62:2026rwr}
\begin{eqnarray}
    {\cal B}(K^+\to\pi^+\nu\bar\nu) &=& \left(9.6^{+1.9}_{-1.8}\right)\times10^{-11}\,.
\end{eqnarray}
In contrast to the Belle II $B^+$-decay measurement, the updated NA62 result is fully consistent with the SM expectation~\cite{Buras:2022wpw,DAmbrosio:2022kvb}. This places constraints on the flavor structure of any new physics that attempts to accommodate the Belle~II measurement.

Our goal is to investigate whether a minimal extension of the SM with an ALP and a relatively light massive sterile neutrino can provide a consistent framework for understanding the anomalies mentioned above. For an ALP with mass near the $\pi^0$ mass, we consider ALP-$\pi^0$ mixing in $B\to\pi K$. We also consider an ALP-sterile neutrino coupling to explain the branching ratio in Eq.~\eqref{BelleIIanomaly} and examine the distribution of signal events across the available kinematic range. Our model improves on the SM description of this spectrum but does not fully reproduce the enhancement in its intermediate region. Our framework includes the minimum set of free parameters needed to accommodate the anomalies while remaining consistent with the neutral current processes, $K_L\to\pi^0 + {\rm inv}$, $B^+\to K^{*+} + {\rm inv}$ and $B^0\to K^{(*0)} + {\rm inv}$, and with the anomalous magnetic moments of the electron and muon.

The paper is structured as follows. In Section~\ref{sec:alpi}, we present our model, including the new particles and their interactions. In Section~\ref{sec:obs}, we examine the various observables -- anomalies and constraints -- that our model sets out to investigate. In Section~\ref{sec:results}, we demonstrate through fits to the data how our model explains the anomalies. We conclude in Section~\ref{sec:conc}.


\section{ALP interactions}\label{sec:alpi}
\noindent
We consider a minimal extension of the SM with an ALP, $a$, and a sterile neutrino, $\nu_N$. The interactions of $a$ and $\nu_N$ with each other and the SM are described by the flavor-conserving Lagrangian,
\begin{align}
  \label{eq:ALP-Lag}
  {\cal L}_a = \frac{1}{2}(\partial_\mu a)(\partial^\mu a) - \frac{1}{2}m_a^2 a^2
  - \frac{1}{4}\kappa a F^{\mu\nu}\tilde F_{\mu\nu}\nn\\
  - i  \eta_t \frac{m_t}{f_a}\bar t \gamma_5 t a
  + i g_{N}\left({\bar\nu_N}\gamma_5\nu_N\right) a\,,
\end{align}
where $f_a$ is the ALP decay constant, $\eta_t$ is the ALP-top quark coupling, and $g_{N}$ is the coupling between $a$ and $\nu_N$. Note that the Lagrangian contains no direct coupling to active neutrinos. However, mixing between the sterile and active neutrinos generates $a\nu_N{\bar\nu}$ and $a\nu{\bar\nu}$ interactions, where $\nu$ without a subscript denotes an active neutrino. The mixing between the four flavor eigenstates, $\nu_\alpha$ ($\alpha = e,\mu,\tau,N$), and mass eigenstates $\nu_{i}$ ($i = 1,\cdots,4$) is described by a $4\times4$ orthogonal matrix $U$~\cite{Datta:2023iln}:
\begin{equation}
	(\nu_\alpha)_{L/R} = \sum_{i=1}^4 U_{\alpha i} \,(\nu_{i})_{L/R}\,,\label{eq:nu-mix}
\end{equation}
for both left- and right-handed neutrinos. The unitarity of $U$ implies $\sum_\alpha |U_{\alpha 4}|^2 = 1$. 
For simplicity, we work in the two-state $(\nu_\mu,\nu_N)$ mixing limit, i.e., $U_{\ell 4} = 0$ for $\ell=\tau, e$. The above condition then simplifies to $|U_{N 4}|^2 + |U_{\mu 4}|^2 = 1$. We take $\nu_4$ to be heavier than $a$, with mass in the MeV--GeV range, such that it is lighter than a $B$ meson but heavier than a kaon. Then, the $\nu_\mu$--$\nu_N$ mixing is expected to be very small, $|U_{\mu 4}|^2 \ll 1$, and $\nu_N$ is mostly $\nu_4$.

 Two distinctive features of our model are that the interactions in Eq.~\eqref{eq:ALP-Lag} are flavor conserving and that the ALP decays promptly. This differs from other ALP solutions to the $B^+\to K^+\nu{\bar\nu}$ excess where the $b\to s$ flavor-changing neutral current (FCNC) couplings appear at tree level~\cite{Altmannshofer:2023hkn}, and the ALP is long-lived~\cite{Altmannshofer:2023hkn,Alda:2025uwo}.

Following the solution to the $B\to K\pi$ puzzle in Ref.~\cite{Bhattacharya:2021shk}, we assume that the $a$ mixes with the $\pi^0$ and has a mass close to $m_{\pi^0}$. Denoting the mass eigenstates of $a$ and $\pi^0$ by the subscript ``p'', the flavor and mass eigenstates of these two particles are related by
\begin{align}
\begin{pmatrix}
	a \\
	\pi^0 
\end{pmatrix}
= 
\begin{pmatrix}
	\cos\theta & \sin\theta\\
	-\sin\theta &  \cos\theta
\end{pmatrix}
\begin{pmatrix}
	a_\mathrm{p}\\
	\pi^0_\mathrm{p}
\end{pmatrix},
\end{align}
where $\theta$ is the mixing angle.

\begin{figure}[t]
\subfloat[]{\includegraphics[width=0.24\textwidth]{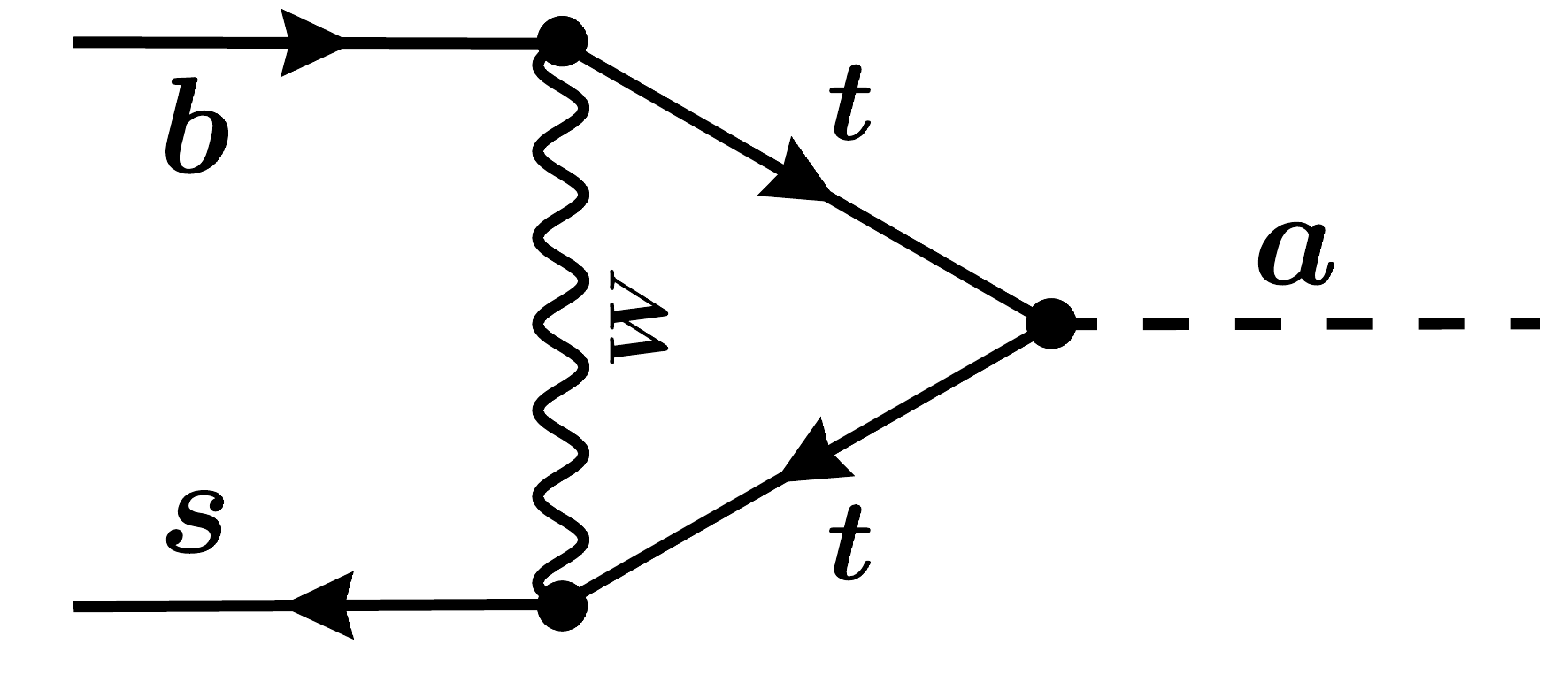}\label{fig:bsa}}
\subfloat[]{\includegraphics[width=0.24\textwidth]{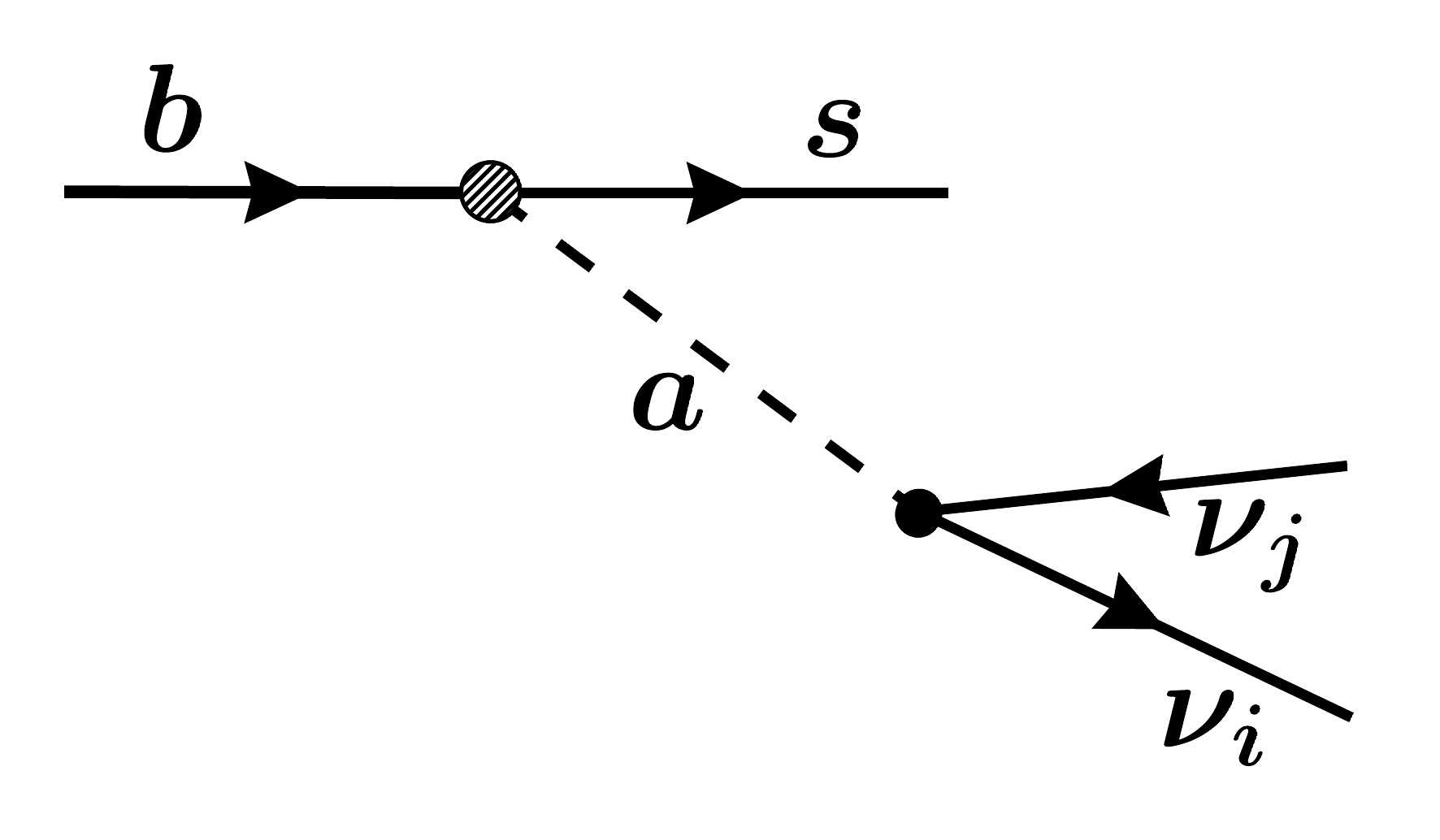}\label{fig:bsnn}}
\caption{Feynman diagrams for (a) the effective $s$-$b$-$a$ vertex, and (b) $b\to s\nu_i\bar{\nu_j}$. }
\label{fig:feyn}
\end{figure}

In this setup, there are two production mechanisms for the ALP. The first is via the penguin processes, e.g., $B\to K a$ and $K\to\pi a$, where the internal top quark is crucial (see Fig.~\ref{fig:bsa}). The relevant FCNC interactions arise only at one loop and involve the top quark and the $W$ boson. These loops simultaneously generate $b \to s$ transitions, mediating $B\to K a$ decays, and $s\to d a$ transitions, enabling the rare kaon decays, $K^+\to\pi^+ a$ and $K_L\to\pi^0 a$. The second production mechanism is through the ALP-$\pi^0$ mixing, whereby the $\pi^0$ produced oscillates to an ALP.

Integrating out the heavy $W$ and top-quark fields gives the effective Lagrangian for the $b\to s$ transition,
\begin{equation}
  \cL_{a}^{bs} = g_{bs} \bar s P_R b\,a\,, \label{eq:Lag_sba}
\end{equation}
where $g_{bs}$ is the effective $\bar{s} b a$ coupling. At the electroweak matching scale, $\mu_\mathrm{EW}$, $g_{bs}$ is given by~\cite{Bhattacharya:2021shk},
\begin{align}\label{eq:g_bsa}
  g_{bs} \simeq \frac{i\sqrt{2} G_F m_b  m_t^2}{16 \pi^2 \,f_a}
  V_{ts}^\ast V^{ }_{tb}\,\eta_t \log\frac{\La^2}{m_t^2}\,,
\end{align}
where $G_F$ is the Fermi constant, $V_{ij}$ are Cabibbo-Kobayashi-Maskawa (CKM) matrix elements, and $\La = 4\pi f_a$ is the new physics cutoff scale.
The effective Lagrangian $\cL_{a}^{sd}$ and the associated coupling $g_{sd}$ corresponding to kaon decays, can be obtained from Eqs.~\eqref{eq:Lag_sba} and~\eqref{eq:g_bsa} by substituting the  quark flavor labels $\{b\}\to\{s\}$ and $\{s\}\to\{d\}$.

%

\section{Observables} \label{sec:obs}

We first examine $B \to  \pi K $ decays, where the ALP mixing with $\pi^0$ provides a resolution to the long-standing puzzle. We then turn to the $B \to K^{(*)} + \mathrm{inv}$ channels, where the same ALP-top coupling generates an effective $b\to s a$ transition that contributes through the off-shell ALP exchange. Finally, we consider constraints from rare kaon decays, which probe the analogous $s\to d a$ transition. We then briefly discuss ALP contributions to lepton anomalous magnetic moments and $b\to s\ell^+\ell^-$ decays. 

\subsection{\texorpdfstring{$\boldsymbol{B \to \pi K}$}{B -> pi K}}

Closely following Ref.~\cite{Bhattacharya:2021shk}, we first review the SM description of $B\to \pi K$ decays before introducing the ALP contribution that addresses the experimental puzzle. Here, we denote the decay amplitude for the process $B^{i + j}\to\pi^i K^j$ as $A^{ij}$, where $i,j = -,0,+$ denote the charges of the particles.
In the SM, the four $B\to\pi K$ amplitudes are related via the isospin quadrilateral,
\begin{align}
	\sqrt{2} A^{00} + A^{-+}  = \sqrt{2} A^{0+} + A^{+0}\,.
\end{align}
These amplitudes can be expressed in terms of six topological diagrams as,
\begin{align}
\label{eq:amps}
A^{+0} &= -P'_{tc} + P'_{uc} e^{i\gamma} -\frac{1}{3}
\pewcp\,, \\
\sqrt{2} A^{0+} &= - T' e^{i\gamma} -C' e^{i\gamma}
+P'_{tc} - P'_{uc} e^{i\gamma} \nl
&\phantom{=}\ - \pewp -\frac{2}{3} \pewcp\,, \\
A^{-+} &= -T' e^{i\gamma} + P'_{tc} -P'_{uc}
e^{i\gamma} -\frac{2}{3} \pewcp\,, \\
\sqrt{2} A^{00} &= -C' e^{i\gamma} - P'_{tc} + P'_{uc} e^{i\gamma} \nl
&\phantom{=}\ - \pewp -\frac{1}{3} \pewcp\,,
\end{align}
where $P'_{tc}$ and $P'_{uc}$ are QCD penguins, $\pewp$ and $\pewcp$ are the color-allowed and color-suppressed electroweak penguins (EWPs), and $T'$ and $C'$ are the color-favored and color-suppressed tree amplitudes. We neglect annihilation topologies, which are expected to be very small. A detailed discussion of these diagrammatic contributions can be found in Ref.~\cite{Baek:2004rp}.

These relations can be further simplified by imposing SU(3) flavor symmetry. Under this symmetry, EWP and tree amplitudes are related via,
\begin{align}
	\pewp + \pewcp
	&=
	\frac{3}{2} \frac{c_9 + c_{10}}{c_1 + c_2} R (T' + C')\,,\label{eq:SU3-F-a}\\
	\pewp - \pewcp
	&=
	\frac{3}{2} \frac{c_9 - c_{10}}{c_1 - c_2} R (T' - C')\,,\label{eq:SU3-F-b}
\end{align}  
where $c_i$ are the $\Delta B = 1$ Wilson 
coefficients~\cite{Buchalla:1995vs}, and $R$ is the magnitude of the ratio of CKM matrix elements, $R = |V^*_{tb} V^{}_{ts}/V^*_{ub} V^{}_{us}|$.

If instead, the $SU(2)$ isospin symmetry exhibited by $B\to \pi K$ decays is imposed, the EWP-tree relations are simpler\footnote{Note that both sets of EWP-tree relations are obtained by neglecting $c_7$ and $c_8$, which are an order of magnitude smaller than $c_9$ and 
$c_{10}$~\cite{Buchalla:1995vs}.}~\cite{Bhattacharya:2025rrv}:
\begin{align}
	\pewp
	&=
	\frac{3}{2} \frac{c_9 + c_{10}}{c_1 + c_2} R \, C'\,,\label{eq:SU2-I-a}\\
	\pewcp
	&=
	\frac{3}{2} \frac{c_9 + c_{10}}{c_1 + c_2} R \, T'\,.\label{eq:SU2-I-b}
\end{align}

Currently, $B \to \pi K$ data cannot be \emph{fully} explained by the SM amplitudes in Eq.~\eqref{eq:amps}. For example, keeping only the leading diagrams (up to  $\mathcal{O}(\lambda)$, where $\lambda$ is the Wolfenstein parameter), it can be shown that the direct CP-asymmetries in $B^+ \to \pi^0 K^+$ and $B^0 \to \pi^- K^+$ should be approximately equal~\cite{Bhattacharya:2021shk}.\footnote{In Refs.~\cite{Gronau:2005kz,Bhattacharya:2025rrv} it was shown that the branching ratios and direct CP asymmetries in the four $B\to\pi K$ decays follow a precise symmetry-based relationship that does not require assumptions about sizes of topological diagrams. Experiments have found that the data obey this relationship within current uncertainties~\cite{Belle-II:2023ksq}. Still, fits to $B\to\pi K$ data reveal that a consistent picture requires non-leading diagrams to be somewhat enhanced.} Experimentally, however, these asymmetries are not only unequal but also have opposite signs~\cite{Bhattacharya:2021shk}. A comprehensive fit including all six diagram topologies likewise fails to describe the data, as we show in Section~\ref{sec:results}. 

To resolve this tension, Ref.~\cite{Bhattacharya:2021shk} proposed a model in which an ALP mixes with $\pi^0$. We now summarize the key elements of this solution and update the analysis using the latest experimental measurements.

The interaction in Eq.~\eqref{eq:Lag_sba} enables ALP production via $B\to aK$. To contribute to $B\to \pi K $ decay, $m_a \sim m_{\pi^0}$ so that the ALP decays promptly to two photons within the detector volume. This occurs either through a direct coupling to photons or via $a$--$\pi^0$ mixing, followed by the SM decay $\pi^0\to 2\gamma$. The general form of the effective ALP-photon coupling is
\begin{align}\label{eq:a2gamgam}
  g_{a\gamma\gamma} =  \sin\theta\, g_{\pi\gamma\gamma} + \kappa_\text{eff}\,,
\end{align}
where $\kappa_\text{eff}$ contains both the tree-level ($\kappa$ in Eq.~\ref{eq:ALP-Lag}) and a top-quark loop contribution~\cite{Bauer:2017ris} which is negligible in our case. We assume that  the  ALP-photon coupling is dominated by the $a$--$\pi^0$ mixing term. The ALP contribution to the $B\to  a K \to \pi K$ amplitude is then
\begin{align}\label{eq:B2kpi_Anp}
  \mathcal{A} = g_{bs} \langle K|  \bar{s} P_R b| B \rangle \sin\theta 
  = g_{bs} \sin\theta \, \frac{m_B^2 - m_K^2}{2 (m_b - m_s)}f_0(q^2)\,,
\end{align}%
where the $B\to K$ matrix element is expressed in terms of the form factor $f_0(q^2)$, evaluated at $q^2 = m_a^2$~\cite{Gubernari:2023puw}. Following Ref.~\cite{Bhattacharya:2021shk} we take $\sin \theta = 0.1$. The ALP contribution $\mathcal{A}$ modifies the amplitudes for $B^+\to \pi^0K^+$ and $B_d^0\to \pi^0K^0$ decays, corresponding to the substitutions,
\begin{align}
	\sqrt{2} A^{0+} \to \sqrt{2} A^{0+} - \mathcal{A}\,,
	 \quad \sqrt{2} A^{00} \to \sqrt{2} A^{00} - \mathcal{A}\,,
\end{align}
in Eq.~\eqref{eq:amps}. The magnitude of $\mathcal{A}$ is determined from a $\chi^2$ fit to $B\to \pi K$ data.  We  perform a fit using the latest experimental data, presenting results for both the SU(3) flavor symmetry and SU(2) isospin symmetry assumptions. 

An off-shell contribution to $B \to \pi K$ via  $b \to s a \to s q \bar q$ (with $q=u,
d$) that depends on the difference $\eta_u - \eta_d$ of the ALP couplings to light quarks (similar to the top quark coupling in Eq.~\ref{eq:ALP-Lag}) modifies $\sin\theta$ as follows~\cite{Bhattacharya:2021shk}:
\begin{align}
\label{eq:eff-mixing}
	\sin \theta
	\rightarrow
	\sin\theta
	+
	\frac{m^2_{\pi^0}}{m^2_{\pi^0} - m^2_a}
	\frac{\eta_u - \eta_d}{2 \sqrt{2}}
	\frac{f_\pi}{f_a}\,.
\end{align}
The couplings $\eta_u$ and $\eta_d$ do not appear in 
Eq.~(\ref{eq:ALP-Lag}), but
renormalization-group evolution (RGE) of $\eta_t$ generates $\eta_u \simeq -0.1 \eta_t$ and $\eta_d \simeq 0.1
\eta_t$ for $f_a=1$~TeV~\cite{Bauer:2020jbp}, giving $\eta_u - \eta_d \simeq -0.2 \eta_t$. For an ALP--$\pi^0$ mass difference of
$1~\mathrm{keV}$ and typical values of $\eta_t$ and $f_a$, the second term in
Eq.~\eqref{eq:eff-mixing} (the off-shell contribution) is about $10^{-2}$, and more suppressed for larger mass splittings. The mixing is therefore dominated by the on-shell contribution, and $\sin\theta \simeq 0.1$.  
Note that the ALP lifetime is $\tau_{\pi^0}/\sin^2\theta \sim 10$~femtosecond.

\subsection{
  \texorpdfstring{$\boldsymbol{B \to K^{(\ast)} + \mathrm{inv}}$}{B->K(*) nu nu}}

The $\bar{s} b a$ coupling in Eq.~\eqref{eq:Lag_sba}, together with the ALP coupling to
sterile neutrinos $g_N$, mediates a tree-level $b\to s \nu_4 \bar \nu_4$ transition via an
ALP mediator, as in Fig.~\ref{fig:bsnn}. If $m_a > 2 m_{\nu_4}$, the ALP can decay on-shell to $\nu_4 \bar \nu_4$, increasing its total width and reducing $\mathcal{B}(a \to 2\gamma)$. This undermines the $B \to \pi K$ solution discussed above. We therefore require $m_a < 2 m_{\nu_4}$, ensuring that $b\to s \nu_4 \bar \nu_4$ proceeds through an off-shell ALP. The corresponding effective Hamiltonian is
\begin{align}\label{eq:Lag_b2sNN}
  \mathcal{H}_\mathrm{eff}
  =
    C_{bs}(q^2)
    (\bar s P_Rb)(\bar \nu_4  \gamma_5 \nu_4 )\,,
  \end{align}
with
\begin{align}
    C_{bs}(q^2) = \frac{ig_{bs} \,g_{N} |U_{N4}|^2}{q^2 - m_a^2 + i m_a \Gamma_a}\,,
\end{align}
where $q^2$ is the invariant mass of the sterile neutrino pair and $\Gamma_a$ is the
total ALP decay width. This operator contributes to both $B \to K \nu_4 \bar \nu_4$ and
$B \to K^{\ast} \nu_4 \bar \nu_4$ decays.

In terms of $\beta = \sqrt{(1-4 m_{\nu_4}^2/q^2)}$, $\Delta_{BM} = m_{B}^2 - m_{M}^2$, and the K\"all\'en function
$\lambda_{BM}\equiv \lambda(m_B^2, m_{M}^2, q^2)$, the
differential decay width for $B\to K \nu_4 \bar\nu_4$ is
\begin{align}
\label{eq:width-BKNN}
  \frac{\mathrm{d}\Gamma (B\to K \nu_4 \bar \nu_4)}{\mathrm{d}q^2}
  =  \frac{\beta\, \lambda_{BK}^{1/2}}{512 \pi^3 m_B^3}
 \frac{\Delta_{BK}^2 \,q^2 f_0^2}{(m_{b} - m_{s})^2} 
  |C_{bs}|^2\,,
\end{align}
and for $B\to K^{\ast}\nu_4 \bar \nu_4 $,
\begin{align}
\label{eq:width-BKstNN}
  \frac{\mathrm{d}\Gamma (B \to K^{\ast} \nu_4 \bar \nu_4)}{\mathrm{d}q^2}
  =
  \frac{\beta\,\lambda_{BK^{\ast}}^{3/2}}{512 \pi^3 m_B^3}
  \frac{ q^2 A_0^2 }{(m_{b}+m_{s})^2} |C_{bs}|^2,
\end{align}
where for brevity we have suppressed the $q^2$ dependence in $C_{bs}(q^2)$  and the $B\to K^{(\ast)}$ form factors $f_0(q^2)$ and $A_0(q^2)$~\cite{Gubernari:2023puw,Bharucha:2015bzk}.

Additionally, neutrino mixing-induced $a \nu_N\bar \nu$
and $a \nu\bar \nu$ interactions together with the $g_{bs}$ coupling give rise to  $b \to s \nu_4 \bar \nu$ and
$b \to s \nu\bar \nu$ transitions, respectively. The $b \to s \nu_4 \bar \nu$ transition is
mediated by an off-shell ALP, with contributions to $B \to K^{(\ast)} + \mathrm{inv}$ decay widths analogous to Eqs.~\eqref{eq:width-BKNN} and \eqref{eq:width-BKstNN}, differing in factors related to neutrino-mixing 
and the phase space, and equivalent to replacing $\beta \rightarrow (1 - m_{\nu_4}^2/q^2)^2$ and $|U_{N4}|^4 \rightarrow |U_{N4}|^2 (1-|U_{N4}|^2)$. On the other hand, the $b \to s \nu\bar \nu$ transition can proceed via resonant
on-shell ALP decay, $b\to s a$ followed by $a \to \nu\bar \nu$. The corresponding contributions to the branching ratios are calculated in the narrow-width approximation:
\begin{align}
\label{eq:B2Knunu_onshell}
  \mathcal{B}(B\to K^{(\ast)} \nu\bar\nu)
  \simeq
  \mathcal{B}(B\to K^{(\ast)} a) \frac{\Gamma(a \to \nu\bar\nu)}{\Gamma_a}\,,
\end{align}
where the partial width for $a\to \nu\bar\nu$ is~\cite{Datta:2023iln}
\begin{align}
  \Gamma(a\to \nu\bar\nu)
  =
  \frac{|\gn|^2 }{8\pi} (1-|U_{N 4}|^2)^2\, m_a\,,
\end{align}
and the partial widths for the two-body decays $B\to K^{(\ast)} a$ are
\begin{align}\label{eq:B2Ka_rate}
  \Gamma(B \to K a)
  &=
    \frac{|g_{bs}|^2 }{64\pi\,m_B^3}
    \left(\frac{m_B^2- m_K^2}{m_b-m_s}\right)^2
    f_0^2 \,\lambda^{1/2}_{BK}\,,\\
  \Gamma(B \to K^\ast a)
  &=
    \frac{|g_{bs}|^2 }{64\pi\, m_B^3\,(m_b+m_s)^2}
    A_0^2 \,\lambda^{3/2}_{BK^\ast}\,,
\end{align}
where $f_0$, $A_0$ and $\lambda_{BK^{(\ast)}}$ are evaluated at $q^2=m_a^2$.

The total ALP contribution 
to $\mathcal{B}(B \to K^{(\ast)} + \mathrm{inv})$, 
\begin{align}
  \label{eq:total-br-alp}
  &\mathcal{B}(B \to  K^{(\ast)} + \mathrm{inv})_{\mathrm{ALP}}\notag\\
  &=\mathcal{B}(B \to  K^{(\ast)} \nu_4  \bar \nu_4)
  + 2\, \mathcal{B}(B \to  K^{(\ast)} \nu_4  \bar \nu)\notag\\
    &+ \mathcal{B}(B \to  K^{(\ast)} a, a \to \nu\bar \nu)\,,
\end{align}
 is dominated by the $b\to s \nu_4 \bar \nu_4$ term. Although decays with sterile-active and active-active neutrinos in the final state have larger phase space than $b \to s \nu_4 \bar \nu_4$, their contributions are negligible due to the small $\nu$--$\nu_N$ mixing in the parameter space of interest. This mixing is constrained by rare kaon decays, as we discuss in the next subsection. In Eq.~\eqref{eq:total-br-alp}, we only include the on-shell ALP contribution from direct $B \to K^{(\ast)} a$ production followed by $a \to \nu\bar \nu$. An additional contribution from ALP--pion mixing ($B \to K^{(\ast)}\pi^{0} \to K^{(\ast)} a$), followed by $a\to \nu\bar \nu$ is doubly suppressed by the small neutrino mixing and the $a$--$\pi^0$ mixing.

The interaction in Eq.~\eqref{eq:Lag_b2sNN} also modifies the invisible decay $B_s \to \mathrm{inv}$. Currently, there is no experimental measurement of this process, but recently an upper bound of $5.6 \times 10^{-4}$ $(90\%~\mathrm{CL})$ on the branching ratio was obtained~\cite{Alonso-Alvarez:2023mgc} from ALEPH data~\cite{ALEPH:2000vvi}. The SM expectation for the branching ratio is $\sim 5 \times 10^{-15}$~\cite{Bhattacharya:2018msv}.  In our model, the new physics contribution is again dominated by $B_s \to \nu_4\nu_4$, with branching ratio,
\begin{align}
	\mathcal{B}(B_s \to \nu_4 \nu_4) = \frac{\beta\,f_{B_s}^2 m_{B_s}^5}{32\pi (m_b+m_s)^2} |C_{bs}|^2\,,
\end{align}
where  $\beta$ and $C_{bs}$ are evaluated at $q^2=m_{B_s}^2$, and $f_{B_s}$ is the $B_s$ decay constant.

\subsection{Rare kaon decays and other constraints}
As in the case of $B$ decays, the ALP induces transitions $s \to d + \mathrm{inv}$ (with $\mathrm{inv} = \nu_4\bar \nu_4$, $\nu_4 \bar \nu$, or $\nu\bar\nu$), modifying the SM decay rates of the rare kaon decays $K_L\to\pi^0 \nu\bar\nu$ and $K^+\to\pi^+ \nu\bar\nu$. These decays are highly suppressed in the SM and can be calculated with excellent precision, making their measurements~\cite{NA62:2026rwr,Redeker:2024joz} powerful probes of new physics. In our model, $K \to \pi \nu_4 \bar \nu_4$ and $K \to \pi \nu_4 \bar \nu$ are kinematically forbidden if $m_{\nu_4} > m_K - m_\pi$. Because the on-shell ALP contribution $K\to \pi a$, $a\to \nu\bar\nu$, depends on $\nu$--$\nu_N$ mixing, the latter can be constrained by rare kaon decays.

The NA62 experiment~\cite{NA62:2021zjw} has set stringent limits on $\mathcal{B}(K^+\to \pi^+ X)$ 
for a feebly-interacting scalar boson $X$ with mass $m_X \in [0, 110]$~MeV and $[160, 260]$~MeV. Since $m_a \simeq m_{\pi^0}$ in our scenario, these bounds do not apply. 
However, the KOTO experiment  has searched for the decay  $K_L\to \pi^0 X$, where $X$ is an invisible boson. For $m_X=135~\mathrm{MeV}$, the 90\% CL upper bound is $\mathcal{B}(K_L \to \pi^0X) \le  1.6 \times 10^{-9}$~\cite{KOTO:2024zbl}, which our model must satisfy. In the narrow-width approximation, the on-shell ALP contribution is
\begin{align}
  \mathcal{B}(K_L\to \pi^0+\mathrm{inv})
  \simeq
  \mathcal{B}(K_L\to \pi^0 a) \frac{\Gamma(a \to \nu\bar\nu)}{\Gamma_a}\,.
\end{align}

The branching ratio for $K_L\to \pi^0 a$ is given by the standard two-body decay formula~\cite{pdg},
\begin{align}
  \mathcal{B}(K_L\to \pi^0 a)
  =
  \tau_{K_L}\frac{|\vec{p}_f|}{8\pi m^2_{K_L}} |\mathcal{\tilde A}|^2,
\end{align}
where $|\vec{p}_f|$ is the momentum of the final-state particles and $\mathcal{\tilde A}$ is the amplitude of $K_L\to \pi^0 a$.

The $K_L\to \pi^0 a$ amplitude receives two contributions. The first arises from the penguin-loop-induced $s \to d \,a$ transition discussed above, which produces the ALP directly in kaon decay. The second comes from $a$--$\pi^0$ mixing via $K_L\to \pi^0\pi^0 \to \pi^0 a$, where $K_L\to \pi^0\pi^0$ is a purely SM decay. We denote these amplitudes as $\mathcal{\tilde A}_\text{direct}$ and $\mathcal{\tilde A}_\text{mix}$, respectively.

$\mathcal{\tilde A}_\text{direct}$ is approximately,
\begin{align}
  \mathcal{\tilde A}_\text{direct}
  \simeq 
  \operatorname{Re}(g_{sd}) \frac{m_K^2 - m_\pi^2}{2(m_s - m_d)} f^K_0(q^2)\,,
\end{align}
where $f^K_0(q^2)$ is the $K \to \pi$ form factor~\cite{Bernard:2009zm,FlaviaNetWorkingGrouponKaonDecays:2010lot} evaluated at $q^2=m_a^2$, giving $f^K_0(m_a^2)\simeq 1$.

To calculate $\mathcal{\tilde A}_\text{mix}$, we first determine the SM amplitude for
$K_L\to \pi^0\pi^0$ using lattice results for the isospin amplitudes for $K^0 \to (\pi\pi)_{I}$ 
from the RBC-UKQCD collaboration~\cite{RBC:2015gro,RBC:2020kdj}. Denoting the
$I=0, 2$ isospin amplitudes as $A_0$ and $A_2$, with corresponding phase shifts
$\delta_I$, the amplitude for $K^0 \to \pi^0\pi^0$ is
\begin{align}
	A(K^0 \to \pi^0\pi^0)
	=
	\sqrt{\frac{2}{3}}
	\left(A_0 e^{i \delta_0} - \sqrt{2} A_2 e^{i\delta_2}
	\right)\,.
\end{align}
%
Using $|K_L\rangle= p_K |K^0\rangle - q_K |\bar{K}^0\rangle$, where $p_K$ and $q_K$ are the usual kaon mixing coefficients, we obtain
$$
A(K_L\to \pi^0\pi^0)= p_K A(K^0 \to \pi^0\pi^0) - q_K A(\bar{K}^0 \to \pi^0\pi^0)\,.
$$
The amplitude for $K_L\to \pi^0\pi^0 \to \pi^0 a$ is then
\begin{align}
	 \mathcal{\tilde A}_\text{mix} = - 2 \sin \theta\, A(K_L\to \pi^0\pi^0)\,,
\end{align}
where the factor of $2$ accounts for either of the two $\pi^0$ mesons mixing with the ALP.  

The total amplitude is the coherent sum $\mathcal{\tilde A} = \mathcal{\tilde A}_\text{direct} + \mathcal{\tilde A}_\text{mix}$, where $\mathcal{\tilde A}_\text{direct}$ depends on the phase $\phi_t$ of the ALP–top coupling $\eta_t$ and $\mathcal{\tilde A}_\text{mix}$ on the $a$--$\pi^0$ mixing angle.

Beyond kaon decay, there are other constraints that are sensitive to the induced ALP couplings to charged leptons. Although the Lagrangian contains no tree-level couplings to leptons, RGE of $\eta_t$ generates $\eta_{\ell} \simeq 0.1 \eta_t$ for $f_a=1~\mathrm{TeV}$~\cite{Bauer:2020jbp}. These couplings induce anomalous magnetic moments of charged leptons, $a_\ell = (g-2)_\ell/2$ at one loop. The corresponding ALP contribution is~\cite{Bauer:2019gfk} 
\begin{align}
  \Delta a_\ell = - \frac{m_\ell^2 \eta_\ell^2}{16 \pi^2 f_a^2}\left[ h_1(x) + \frac{2\alpha}{\pi}\frac{c_{\gamma\gamma}^\text{eff}}{\eta_\ell} \left(\log \frac{\Lambda^2}{m_\mu^2} - h_2(x)\right)\right]\,,
\end{align}
where $x=m_a^2/m_\mu^2 - i\epsilon$, $c_{\gamma\gamma}^\text{eff} = - (\pi f_a /\alpha)\, g_{a\gamma\gamma}$, and the loop functions $h_{1}(x)$ and $h_{2}(x)$ are given in Ref.~\cite{Bauer:2017ris}. Since RGE-induced couplings $\eta_\ell$ are universal, we have $\Delta a_e/\Delta a_\mu = m_e^2 /m_\mu^2$.

The lepton coupling $\eta_e$ together with the $\bar s b a$ coupling also induces the $b \to s e^+ e^-$ transition, which modifies the branching ratios of $B_s \to e^+e^-$ and $B \to K^{(\ast)} e^+ e^-$.  The correction to the former is given by~\cite{Bauer:2021mvw}
\begin{align}
\label{eq:Bs-ee}
  \frac{\mathcal{B}(B_s\to e^+e^-)}{\mathcal{B}(B_s\to e^+e^-)_\text{SM}} = \left| 1 - \Delta_e\right|^2\,,
\end{align}
where $\Delta_e$ reads
\begin{align}
  \Delta_e = \frac{-i \pi v^2 \,\eta_{e}\, g_{bs}}{\alpha V_{ts}^\ast V_{tb} C_{10}^\text{SM}(1- m_a^2/m_{B_s}^2)m_bf_a}\,.
\end{align}
with $C_{10}^\text{SM} \simeq -4.2$ in the SM. 

The branching ratio for $B \to K^{(\ast)} e^+e^-$ with an on-shell ALP in the narrow-width approximation follows analogously from Eq.~\eqref{eq:B2Knunu_onshell}, with the decay $a \to \nu\bar\nu$ replaced by $a \to e^+e^-$. Neglecting terms $\sim m_e^2/m_a^2$, the ALP partial width to electrons is
\begin{align}
	\Gamma(a \to e^+e^-) \simeq  \frac{|\eta_e|^2 m_e^2 m_a}{8 \pi f_a^2}\,.
\end{align}

\section{Results and Discussion}
\label{sec:results}
\begin{table}[t]
  \centering
  \small
  \setlength{\tabcolsep}{-2.5pt}
  \begin{tabular}{@{} c c c c @{}}
    \toprule
     \multirow{2}[3]{*}{Parameter}
      & \multicolumn{2}{c}{$\mathrm{SU(3)}$ flavor}  
      & \multicolumn{1}{c}{$\mathrm{SU(2)}$ isospin}\\
      \cmidrule(l{1em}r{2em}){2-3}\cmidrule(lr){4-4}
      &\multicolumn{1}{c}{Case I} &{Case II}
      &\multicolumn{1}{c}{Case III} \\
    \midrule
    $|T^\prime|$ & $2.34 \pm 0.43$  & $99.10 \pm 2.94$ & $2.34 \pm 0.43$ \\
    $|P^{\prime}_{tc}|$ & $49.33 \pm 0.39$  & $35.48 \pm 2.48$ & $49.35 \pm 0.39$ \\
    $\delta_{Ptc^\prime}$ & $(-80.81 \pm 2.64)^\circ$  & $(-1.42 \pm 0.39)^\circ$ & $(-81.67 \pm 2.57)^\circ$ \\
    $\delta_{C^\prime}$ & $(-89.50 \pm 120.62)^\circ$ & $(-6.22 \pm 0.84)^\circ$ & $(-90.01 \pm 120.94)^\circ$ \\
    $|P^{\prime}_{uc}|$ & $3.49 \pm 1.00$  & $54.28 \pm 0.99$ & $3.41 \pm 0.98$ \\
    $\delta_{Puc^\prime}$ & $(-84.02 \pm 7.37)^\circ$  & $(179.48 \pm 0.26)^\circ$ & $(-85.66 \pm 7.79)^\circ$ \\
    $|\eta_t| \times 10^2$ & $3.09 \pm 0.45$ & $9.91 \pm 0.23$ & $3.22 \pm 0.44$ \\
    $\phi_{t}$ & $(95.97 \pm 4.20)^\circ$ & $(-127.91 \pm 2.50)^\circ$ &$(96.06 \pm 4.36)^\circ$ \\
    \midrule
    $\chi^2_{\min}/\text{dof}$ & 1.42/1 & 0.08/1 & 1.41/1 \\
    p-value & $0.23$ & $0.78$ & $0.23$ \\
    \bottomrule
  \end{tabular}
  \caption{Results of a fit to $B\to \pi K $ data, assuming
    $|C'|/|T'| = 0.2$. The amplitudes are in units of eV. 
    With only the SM amplitudes, Cases I and II  give $\chi^2_{\min}/\text{dof}=16.96/3$, and Case III gives $\chi^2_{\min}/\text{dof}=37.20/3$. }
    \label{tab:fits}
\end{table}

In Table \ref{tab:fits}, we present the results of our $\chi^2$ fit to the latest $B\to\pi K$ data. We use the MINUIT package~\cite{iminuit,James:1975dr} for minimization of the $\chi^2$ function and obtain best-fit values, including their uncertainties. We consider three scenarios with different symmetry assumptions. Cases~I and~II impose SU(3) flavor symmetry with the amplitude relations of Eqs.~\eqref{eq:SU3-F-a} and~\eqref{eq:SU3-F-b}. Motivated by QCD factorization we set $|C'|/|T'|=0.2$~\cite{Beneke:2009ek}.  In Case~I, we constrain the color-favored tree amplitude $|T'|$ to be smaller than the QCD penguin $|P'_{tc}|$, as is typically 
expected~\cite{Baek:2004rp}, and in Case~II we relax this constraint. Case~III only employs SU(2) isospin symmetry, with the relations in Eqs.~\eqref{eq:SU2-I-a} 
and~\eqref{eq:SU2-I-b}. 

The SM-only fit yields a very poor description of the data
as quantified by the minimum $\chi^2$ per degree of freedom (dof): $\chi^2_{\min}/\mathrm{dof} = 16.96/3$ (p-value $=7\times 10^{-4}$) for 
Cases~I and~II, and $\chi^2_{\min}/\mathrm{dof} = 37.20/3$ (p-value $=4.2\times 10^{-8}$) for Case~III. Inclusion of the ALP amplitude significantly improves the fit in all cases. 
Cases~I and~III give good fits with $\chi^2_{\min}/\mathrm{dof} \sim 1.41/1$ (p-value $=0.23$), while Case~II provides an excellent fit with $\chi^2_{\min}/\mathrm{dof} = 0.08/1$ (p-value $=0.78$). 
In Ref.~\cite{Altmannshofer:2024kxb}, the same $B\to\pi K$ observables are fit
with a different amplitude normalization and CKM convention, and $C'/T'$ is allowed
to float. Under the same assumptions and CKM convention, we
reproduce their SM result. Our SM fit differs because we fix $|C'|/|T'|=0.2$.

The three solutions differ in the extracted amplitudes. Cases~I and~III favor a small tree amplitude $|T^\prime| \simeq 2.34$~eV with a large QCD penguin $|P^{\prime}_{tc}| \simeq 49.3$~eV, while Case~II exhibits the opposite hierarchy with $|T^\prime| \simeq 99.1$~eV and $|P^{\prime}_{tc}| \simeq 35.5$~eV. The ALP coupling $|\eta_t|$ varies accordingly, ranging from $\sim 0.03$ in Cases I and III to $\sim 0.1$ in Case~II. These fits determine both the magnitude and phase of $\eta_t$ for the chosen value of $\sin\theta$. In what follows we focus on the commonly studied SU(3) flavor symmetry, and although Case~II is an excellent fit, we do not consider it further because the hierarchy of amplitudes is unexpected.

Table~\ref{tab:btokpi} shows the expected values for the $B \to \pi K$ observables for the best-fit parameters of Case~I. The values agree well with experimental measurements. The largest deviation appears in $A_{CP}(B^0 \to \pi^0 K^0)$, which lies $\sim 1.1\sigma$ from its measured value and provides the dominant $\chi^2$ contribution of 1.30.

\begin{table}
  \centering
  \small
  \begin{tabular}{@{} l @{\hspace{0.1em}} r r r @{\hspace{1em}} r @{}}
    \toprule
  Observable &  \multicolumn{1}{c}{Measurement}  & \multicolumn{1}{r}{Theory}& Pull & \multicolumn{1}{c}{$\chi^2$} \\
    \midrule
    $\mathcal{B}(B^+ \to \pi^+ K^0)\times 10^5$ & $2.39\pm0.06$ & $2.390$ & $0.005$ & $0.000$ \\
    $\mathcal{B}(B^+ \to \pi^0 K^+) \times 10^5$ & $1.32\pm0.04$ & $1.320$ & $-0.001$ & $0.000$ \\
    $\mathcal{B}(B^0 \to \pi^- K^+) \times 10^5$ & $2.00\pm0.04$ & $1.999$ & $0.030$ & $0.001$ \\
    $\mathcal{B}(B^0 \to \pi^0 K^0)\times 10^5$ & $1.01\pm0.04$ & $1.013$ & $-0.064$ & $0.004$ \\
    $A_{CP}(B^+ \to \pi^+ K^0)$ & $-0.003\pm0.015$ & $-0.007$ & $0.268$ & $0.072$ \\
    $A_{CP}(B^+ \to \pi^0 K^+)$ & $0.027\pm0.012$ & $0.025$ & $0.203$ & $0.041$ \\
    $A_{CP}(B^0 \to \pi^- K^+)$ & $-0.083\pm0.003$ & $-0.083$ & $-0.038$ & $0.001$ \\
    $A_{CP}(B^0 \to \pi^0 K^0)$ & $0.00\pm0.08$ & $-0.091$ & $1.140$ & $1.300$ \\
    $S_{CP}(B^0 \to \pi^0 K^0)$ & $0.64\pm0.13$ & $0.635$ & $0.039$ & $0.002$ \\
    \bottomrule
  \end{tabular}
   \caption{Summary of the latest experimental measurements of the CP-averaged branching ratios, direct CP asymmetries $A_{CP}\equiv [\mathcal{B}(\bar B \to \bar F)- \mathcal{B}(B \to F)]/[\mathcal{B}(\bar B \to \bar F) + \mathcal{B}(B \to F)]$ (with final states $F,\bar F$), and
    mixing-induced CP asymmetry $S_{CP}$ of the $B \to K \pi$ modes
    in our analysis~\cite{
    pdg,LHCb:2020byh,Belle-II:2023ksq,Belle-II:2023grc}. 
   The theoretical expectations are for the best-fit parameters for Case~I in Table~\ref{tab:fits}. The
    last two columns list the pull with respect to their experimental values and the $\chi^2$ contribution to the fit.  } 
    \label{tab:btokpi}
\end{table}

\begin{figure}[t]
  \centering
  \includegraphics[width=0.48\textwidth]{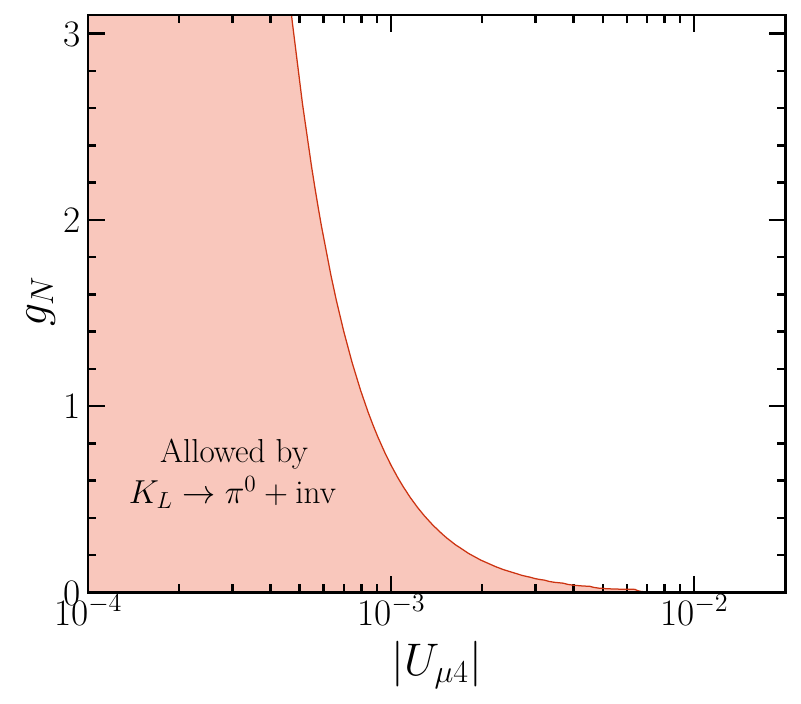}%
  \caption{The 90\%~CL KOTO constraint from $\mathcal{B}(\KLtopinv)$ in the $|U_{\mu4}| - g_{N}$ plane.}%
  \label{fig:koto}
\end{figure}

With $\eta_t$ fixed by the $B \to \pi K$ analysis, we now turn to the $\BtoKinv$ excess. To obtain the SM branching ratios for $B \to K^{(\ast)} \nu \bar \nu$, we use the {\tt flavio} package~\cite{Straub:2018kue}. The $b \to s \nu \nu$ transition depends on $g_N$, $U_{\mu 4}$ and $m_{\nu_4}$. Since $\mathcal{B}(\KLtopinv)$ is independent of $m_{\nu_4}$, the KOTO upper bound directly constrains the $|U_{\mu 4}|$--$g_N$ plane. Figure~\ref{fig:koto} shows the excluded region: for $g_N \sim \mathcal{O}(1)$, we find $|U_{\mu 4}| \lesssim \mathcal{O}(10^{-3})$. We conservatively fix $|U_{\mu 4}| = 5 \times 10^{-4}$ in the analysis below.
%
%
\begin{figure*}
  \centering
  \includegraphics[width=0.31\textwidth]{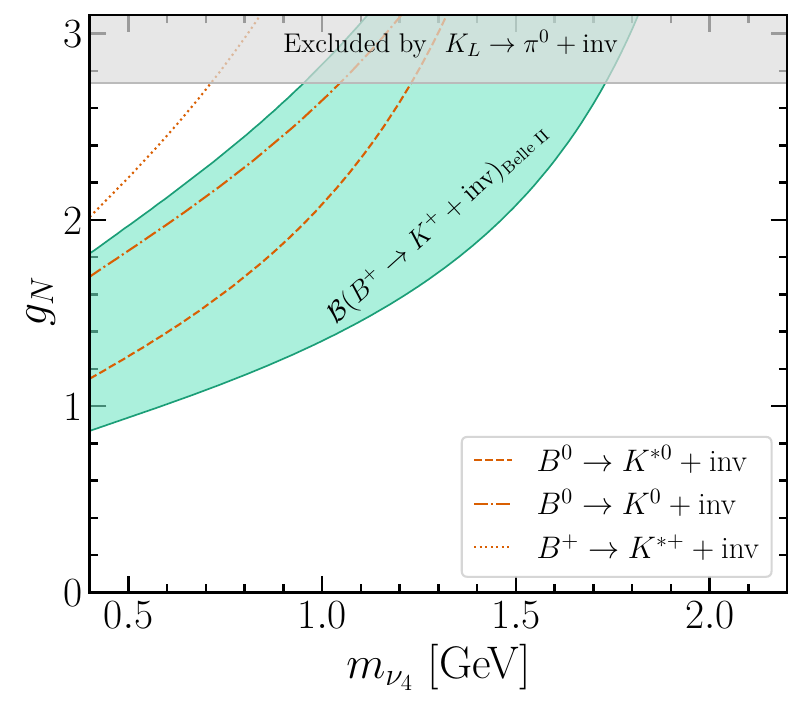}\quad
  \includegraphics[width=0.325\textwidth]{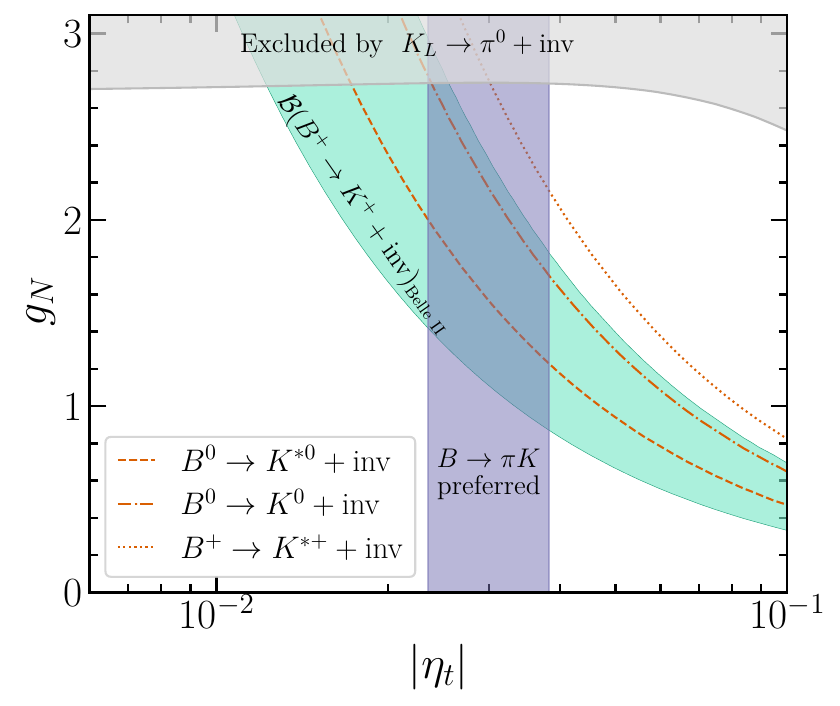}\quad
  \includegraphics[width=0.32\textwidth]{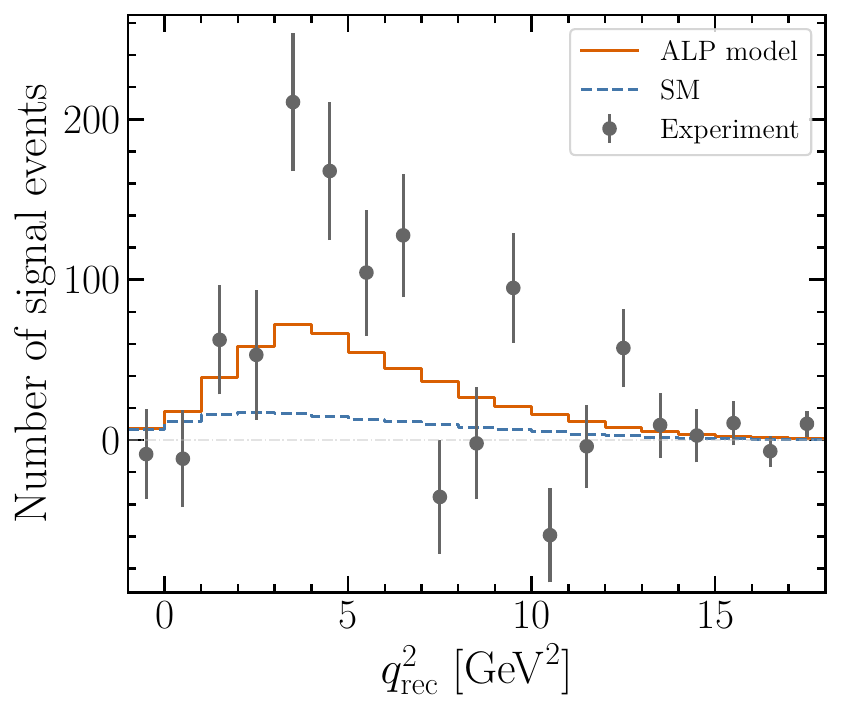}
  \caption{Left panel: Allowed parameter space in the $m_{\nu_4}$--$g_N$ plane for the best-fit $\eta_t$ of Case~I, $\sin\theta=0.1$, $f_a=1$~TeV and $U_{\mu 4} = 5 \times 10^{-4}$.  The green shading corresponds to the 90\% CL region favored by the $\BtoKinv$ excess.  The orange curves show the 90\%~CL upper limits from other $B \to K^{(\ast)}+\mathrm{inv}$ decays. The gray shaded region is ruled out at the 90\%~CL by the non-observation of $\mathcal{B}(\KLtopinv)$ at KOTO. Middle panel: Allowed parameter space in the $|\eta_t|$--$g_N$ plane for $m_{\nu_4}=689$~MeV. The purple shaded region corresponds to the  preferred values of $|\eta_t|$ in Case~I at 90\% CL. The constraint from $B_s\to \mathrm{inv}$ is not shown as it is weaker than the corresponding KOTO constraint. Right panel: Comparison of the background-subtracted $B^+\to K^+ + \mathrm{inv}$ signal spectrum in $q^2_\text{rec}$ bins with the expectation for $m_{\nu_4} = 689~\mathrm{MeV}$ and $g_N=1.38$. The SM spectrum is also shown. }%
  \label{Fig:mN}
\end{figure*}

In Fig.~\ref{Fig:mN} we display the parameter space favored by the $\BtoKinv$ excess. The left panel shows the $m_{\nu_4}$--$g_N$ plane for the best-fit $\eta_t$ of Case~I (see Table~\ref{tab:fits}), $\sin\theta=0.1$, $f_a=1$~TeV and $U_{\mu 4} = 5 \times 10^{-4}$. The green band corresponds to values that explain the Belle~II measurement of $\mathcal{B}(B^+ \to K^+ + \mathrm{inv})$ at $90\%$ CL The orange dashed, dot-dashed, and dotted curves show the $90\%$ CL upper limits from $\mathcal{B}(B^0\to K^{*0}+\mathrm{inv})$, $\mathcal{B}(B^0\to K^0+\mathrm{inv})$, and $\mathcal{B}(B^+\to K^{*+}+\mathrm{inv})$, respectively; the regions above these curves are excluded. The gray band indicates the parameter space ruled out by the KOTO $90\%$ CL bound on $\mathcal{B}(K_L\to \pi^0 + \mathrm{inv})$.

The middle panel shows the allowed region in the $|\eta_t|$--$g_N$ plane for $m_{\nu_4}=689$~MeV, which is the value that best describes the measured spectrum; see below. The purple band marks $|\eta_t|$ values compatible with $B\to \pi K$ data at $90\%$ CL. The overlap of the green and purple regions demonstrates that our model can simultaneously address the $B\to \pi K $ puzzle and the $B^+ \to K^+ \nu\bar\nu$ excess while satisfying all existing constraints.

Having shown that the ALP can explain the $\mathcal{B}(B^+ \to K^+ \nu \bar\nu)$ excess, we now examine whether it can also reproduce the $q^2$ distribution measured by Belle~II. 
We perform a combined $\chi^2$ fit to both the spectrum and the total branching ratio. 

Since the $B^+$ four-momentum is not fully reconstructed, Belle~II reports signal events in {\it reconstructed} $q^2$, defined as $q^2_\text{rec} = m_{B^+}^2 + m_{K^+}^2 - 2 m_{B^+} E^\ast_{K^+}$, with $E^\ast_{K^+}$ the $K^+$ energy in the center-of-mass frame. The expected number of signal events per $q^2_\text{rec}$ bin is~\cite{Bolton:2025fsq}
\begin{align}
	\frac{\mathrm{d} N}{\mathrm{d} q^2_\text{rec}}
	= 
	N_{B}\!\int \!\mathrm{d}q^2 f_{q^2_\text{rec}}(q^2) \epsilon(q^2) \frac{d\mathcal{B}}{\mathrm{d}q^2}\,,
\end{align}
where $N_B= (387 \pm 6) \times 10^6$,  $f_{q^2_\text{rec}}(q^2)$ describes the smearing  of $q^2_\text{rec}$ with respect to $q^2$, $\epsilon(q^2)$ is the signal-selection efficiency~\cite{Belle-II:2023esi}, and $d\mathcal{B}/\mathrm{d}q^2$ is the differential branching ratio including SM contributions. For the signal event counts, we use the background-subtracted values extracted in Ref.~\cite{Bolton:2025fsq}. The smearing function is approximated by the $q^2_\text{rec}$-$q^2$ correlation shown in Fig.~5.1 of Ref.~\cite{Praz:2022bci}.

Fixing $\eta_t$ to the best-fit value of Case~I, we perform a combined $\chi^2$ analysis of the spectrum and total branching ratio. For the fit to the spectrum, we assume Gaussian statistics for bins with more than five events and Poisson statistics for bins with fewer events~\cite{pdg}. The SM fit gives 
 $\chi^2_{\min}/\mathrm{dof} = 94.56/21$ (p-value $=2.6 \times 10^{-11}$), while the fit including the ALP contribution yields $m_{\nu_4} \simeq 689~\mathrm{MeV}$ and $g_{N}\simeq 1.38$, resulting in a much-improved, but still poor fit with 
$\chi^2_{\min}/\mathrm{dof} = 54.73/19 $ (p-value $= 2.56 \times 10^{-5}$). The resulting $q^2_{\rm rec}$ distribution is shown in the right panel of Fig.~\ref{Fig:mN}. Overall, the ALP model provides a significantly better description of the $B^+ \to K^+ + \mathrm{inv}$ spectrum than the SM, although it does not account for the  excess at intermediate $q^2$. 

\begin{table}[t]
  \centering
  \small
  \begin{tabular}{@{} l @{\hspace{1em}} r @{\hspace{1em}} c @{\hspace{1em}} c @{}}
\toprule
\multirow{2}[1]{*}{Observables} & \multicolumn{3}{c}{Branching ratios ($\times 10^{5}$)} \\
 \cmidrule(l{0.5em}r{0em}){2-4}
 &  \multicolumn{1}{c}{Measurement} & SM only &  ALP model\\
\midrule
$\mathcal{B}(B^+ \to K^+ + \mathrm{inv})$ & $2.3 \pm 0.7$~\cite{Belle-II:2023esi}  & $0.47 \pm 0.02$ & $1.58 \pm 0.07$ \\
$\mathcal{B}(B^+ \to K^{*+} + \mathrm{inv})$ & $< 4.0$~\cite{Belle:2013tnz}  & $1.06 \pm 0.10$ & $1.94 \pm 0.19$ \\
$\mathcal{B}(B^0 \to K^0 + \mathrm{inv})$ & $< 2.6$~\cite{Belle:2017oht}  & $0.44 \pm 0.02$ & $1.46 \pm 0.06$ \\
$\mathcal{B}(B^0 \to K^{*0} + \mathrm{inv})$ & $< 1.8$~\cite{Belle:2017oht}  & $0.98 \pm 0.10$ & $1.80 \pm 0.17$ \\
\bottomrule
\end{tabular}
   \caption{Comparison of the measured branching ratios for $B \to K^{(\ast)} +\mathrm{inv}$ decays, with SM and ALP model expectations (with the benchmark values in Fig.~\ref{Fig:mN}). All upper limits correspond to $90\%~\mathrm{CL}$.}
    \label{tab:predictions}
\end{table}

In Table~\ref{tab:predictions} we compare the measured branching ratios for the four $B \to K^{(\ast)} \!+\mathrm{inv}$ with the model predictions obtained with the benchmark values corresponding to the right panel of Fig.~\ref{Fig:mN}. The prediction for the anomalous channel, $\mathcal{B}(B^+ \to K^+ +\mathrm{inv})$, lies roughly within $1\sigma$ of the Belle~II result. The model also predicts enhancements in the other three channels relative to their SM values. In particular, the predicted branching ratio of $B^0 \to K^{*0} +\mathrm{inv}$ coincides with the $90\%~\mathrm{CL}$ experimental upper bound; improved measurements of this decay will provide an important test of the model. The loop that generates the $g_{bs}$ coupling also induces $\bar{d} b a$ coupling, $g_{bd}$, obtained from Eq.~\eqref{eq:g_bsa} by replacing $V_{ts}^\ast$ by $V_{td}^\ast$. For the benchmark point in Fig.~\ref{Fig:mN}, the off-shell contribution gives $\mathcal{B}(B^+\to\pi^+\nu_4\bar\nu_4)=2.9\times10^{-7}$, and the total rate $\mathcal{B}(B^+\to\pi^+ + \mathrm{inv})_{\mathrm{SM+ALP}}=4.3\times10^{-7}$ is below the 90\%~CL bound 
$\mathcal{B}(B^+\to\pi^+\nu\bar\nu)<8.0\times10^{-6}$~\cite{Belle:2017oht}.

Before concluding, we consider consequences of the model for lepton anomalous magnetic moments and $b \to s e^+ e^-$ decays. Using the benchmark values of Fig.~\ref{Fig:mN}, we find $|\Delta a_e| \simeq \mathcal{O}(10^{-15})$, which is well below the current experimental values $|\Delta a_e|_{\mathrm{Rb}} = (0.35 \pm 0.16) \times 10^{-12}$ and ($|\Delta a_e|_{\mathrm{Cs}} = (1.00 \pm 0.26) \times 10^{-12}$~\cite{Fan:2022eto,Aliberti:2025beg}, based on two different measurements of the fine-structure constant using rubidium atoms~\cite{Morel:2020dww} and cesium atoms~\cite{Parker:2018vye}. In contrast, we find a 
non-negligible contribution to the muon anomalous magnetic moment, $\Delta a_\mu \simeq 2.6\times 10^{-11}$, which is consistent with the measured value, $\Delta a_{\mu} = (38 \pm 63) \times 10^{-11}$~\cite{Aliberti:2025beg}. For $b \to s e^+e^-$ decays, we find ALP contributions to be negligible. For $B_s\to e^+e^-$, the ratio in Eq.~\eqref{eq:Bs-ee} differs from unity at the level of $3.6 \times 10^{-5}$; for $B \to K^{(\ast)} e^+e^-$, we find $\mathcal{B}(B^+ \to K^+ e^+ e^-) \sim 5 \times 10^{-14}$ and $\mathcal{B}(B^+ \to K^{\ast +} e^+ e^-) \sim 6 \times 10^{-14}$, which are negligible.

\section{Conclusions} \label{sec:conc}

The larger than expected value of $\mathcal{B}(B^+ \to K^+ +\mathrm{inv})$ measured by the Belle II collaboration has generated a lot of interest recently. Also, a decades old
  puzzle in the penguin dominated decays, $B \to K \pi $  has seen a recent revival in interest. We addressed these two anomalies  with an axion-like particle, $a$ with mass almost the same as that of the $\pi^0$. We calculated the  $ B \to  K a $ penguin generated diagram which contributes to $B \to K \pi^0$ due to the $ a- \pi^0$ mixing and yields a much improved fit to the $ B \to K \pi$ observables relative to the SM. 
  We calculated the ALP's contribution to $\mathcal{B}(B^+ \to K^+ + \mathrm{inv})$ via its off-shell coupling to a $\nu_N\bar{\nu}_N$ pair and made predictions for the other $B \to K^{(*)} + \mathrm{inv} $ modes. We find that a sterile neutrino of mass $\sim 689$~MeV gives a good fit to the Belle~II data, while being consistent with other constraints. However, although $m_{\nu_4} =689$~MeV significantly improves the fit to the $q^2$ distribution relative to the SM, it does not provide a good fit to the spectrum. We also derived constraints on the model from the rare kaon decays, $K^+ \to \pi^+ +\mathrm{inv}$ and
$K_L \to \pi^0 + \mathrm{inv}$. Although we included only the coupling of the ALP to the top quark, RGE effects generate couplings to all leptons. We find tiny ALP contributions to $b \to s e^+ e^-$ and $\Delta a_e$, but an appreciable effect on $\Delta a_\mu$ at a level consistent with data.

\vspace{0.1in}
\acknowledgments
We thank P.~Bolton for providing the Python implementation of $f_{q^2_\text{rec}}$ used in Ref.~\cite{Bolton:2025fsq}, 
and L.~Mukherjee for helpful communications. B.B.~is supported by the U.S.~National Science Foundation through Grant No.~PHY-2310627. A.D.~is supported in part by the U.S.~National Science Foundation under Grant No.~PHY-2309937. G.K.~thanks the support of the U.S. Department of Energy grant DE-SC0024357. D.M.~is supported in part by the U.S.~Department of Energy under Grant No.~de-sc0010504.

\bibliography{refs}

\end{document}